\documentclass[twocolumn]{aastex}
\usepackage{graphics}
\usepackage{tabu}
\usepackage{natbib}
\usepackage{color}
\usepackage{rotating}
\usepackage{multirow}
\usepackage{footnote}
\usepackage{datetime}
\usepackage{amsmath}
\usepackage{mathrsfs}
\usepackage[normalem]{ulem}

\begin{document}
\title{Measuring Turbulence with Young Stars in the Orion Complex} 
\author[0000-0001-6600-2517]{Trung Ha}
\affil{Department of Physics, University of North Texas, Denton, TX 76203, USA}
\email{trungha@my.unt.edu}

\author[0000-0001-5262-6150]{Yuan Li}
\affil{Department of Physics, University of North Texas, Denton, TX 76203, USA}
\affil{Department of Astronomy, University of California, Berkeley, CA 94720, USA}

\author[0000-0002-0458-7828]{Siyao Xu}
\thanks{NASA Hubble Fellow}
\affil{Institute for Advanced Study, 1 Einstein Drive, Princeton, NJ 08540, USA}

\author[0000-0002-5365-1267]{Marina Kounkel}
\affil{Department of Physics and Astronomy, Western Washington University, 516 High St, Bellingham, WA 98225, USA}

\author[0000-0002-1253-2763]{Hui Li}
\thanks{NASA Hubble Fellow}
\affil{Department of Physics, Kavli Institute for Astrophysics and Space Research, Massachusetts Institute of Technology, Cambridge, MA 02139, USA}
\affil{Department of Astronomy, Columbia University, 550 West 120th Street, New York, NY 10027, USA}

\begin{abstract}
Stars form in molecular clouds in the interstellar medium (ISM) with a turbulent kinematic state. Newborn stars therefore should retain the turbulent kinematics of their natal clouds. Gaia DR2 and APOGEE-2 surveys in combination provide three-dimensional (3D) positions and 3D velocities of young stars in the Orion Molecular Cloud Complex. Using the full 6D measurements, we compute the velocity structure functions (VSFs) of the stars in six different groups within the Orion Complex. We find that the motions of stars in all diffuse groups exhibit strong characteristics of turbulence. Their first-order VSFs have a power-law exponent ranging from $\sim0.2-0.5$ on scales of a few to a few tens of pc, generally consistent with Larson's relation. On the other hand, dense star clusters, such as the Orion Nebula Cluster (ONC), have experienced rapid dynamical relaxation, and have lost the memory of the initial turbulent kinematics. The VSFs of several individual groups and the whole Complex all show features supporting local energy injection from supernovae. The measured strength of turbulence depends on the location relative to the supernova epicenters and the formation history of the groups. Our detection of turbulence traced by young stars introduces a new method of probing the turbulent kinematics of the ISM. Unlike previous gas-based studies with only projected measurements accessible to observations, we utilize the full 6D information of stars, presenting a more complete picture of the 3D interstellar turbulence.
\end{abstract}

\section{Introduction}
\label{sec:intro} 
\setcounter{footnote}{0} 
The interstellar medium (ISM) is turbulent. More specifically, the cold dense star-forming molecular clouds are turbulent. For example, \citet{Larson81} finds a power law relation between velocity dispersion and cloud size with an exponent of 0.38, similar to the Kolmogorov law for incompressible turbulence \citep{Kolmogorov41}. Turbulence plays an important role in regulating the cloud dynamics and star formation \citep{2004RvMP...76..125M, Mckee_Ostriker2007,Hennebelle12},
from cloud scales all the way down to the scales much smaller than molecular cores 
\citep{Hul17}.

The ubiquitous interstellar turbulence induces velocity fluctuations on all scales in both diffuse interstellar phases and dense star-forming gas. Statistical measurements of velocity fluctuations in the multi-phase ISM can be done using the Velocity Channel Analysis (VCA), the Velocity Coordinate Spectrum (VCS) \citep{LP00,LP06}, the Delta-variance technique \citep{Stutzki1998,Ossenkopf02}, or the Principal Component Analysis \citep{Heyer04,Roman-Duval11} based on spectroscopic data.
These analyses generally reveal power-law spectra of turbulent velocities \citep{Laz09rev,Chep10}.

Many processes can drive turbulence in the ISM, such as gravity, shear, and various forms of stellar feedback including supernova explosions \citep{Krumholz2016, Pad16}. Turbulent motions shape the filamentary density structures in molecular clouds (e.g., \citealt{Fed16,XuJ19}). Dense cores and protostars subsequently form in dense filaments \citep{And17}.

Besides gas tracers that are commonly used to probe turbulence in the 
volume filling ISM, 
turbulent velocities can also be sampled using point sources 
such as molecular cores formed at the density peaks generated in highly supersonic turbulence. 
For example, \citet{Qi12} developed statistical measurements of 
core-to-core velocity dispersion over a range of length scales , i.e., 
the core velocity dispersion technique. \citet{Qi18} find 
signatures for a turbulent power-law spectrum in the Taurus Molecular Cloud, and demonstrate that the statistics of turbulent velocities are imprinted in the velocities of dense cores \citep{Xu20}. 

Because prestellar cores are nurseries of stars, newborn stars should naturally inherit the velocities of cores and hold the fossil information on the turbulent velocities of their natal clouds, as long as dynamical interactions of young stars have not erased that information.

In this work, we carry out the first statistical analysis of star-to-star velocity structure function (VSF) across the Orion complex over a range of length scales, from $\sim200$ pc down to $<1$ pc. These newly formed stars are expected to still ``remember" the local gas kinematics at their birth, and can be used to recover the turbulent motions in the parent cloud. With direct measurements on 3D velocities and 3D positions, the statistical study of turbulent velocities with stars do not suffer from the 
complex convolution between density and velocity and the projection effect as that with the commonly used gas tracers. 

In Section~\ref{sec:data}, we describe the data used in this analysis and how we compute the VSFs. We present the VSF for each group in the Orion complex in Section~\ref{sec:results}, and discuss how the slope and amplitude is related to the location and history of the stellar system. In Section~\ref{sec:discussions}, we discuss the uncertainties and biases of the analysis, and how our method and results compare with previous works on turbulence in the ISM.

\section{Data Processing}\label{sec:data}

\begin{figure}
    \centering
    \includegraphics[width=0.8\linewidth]{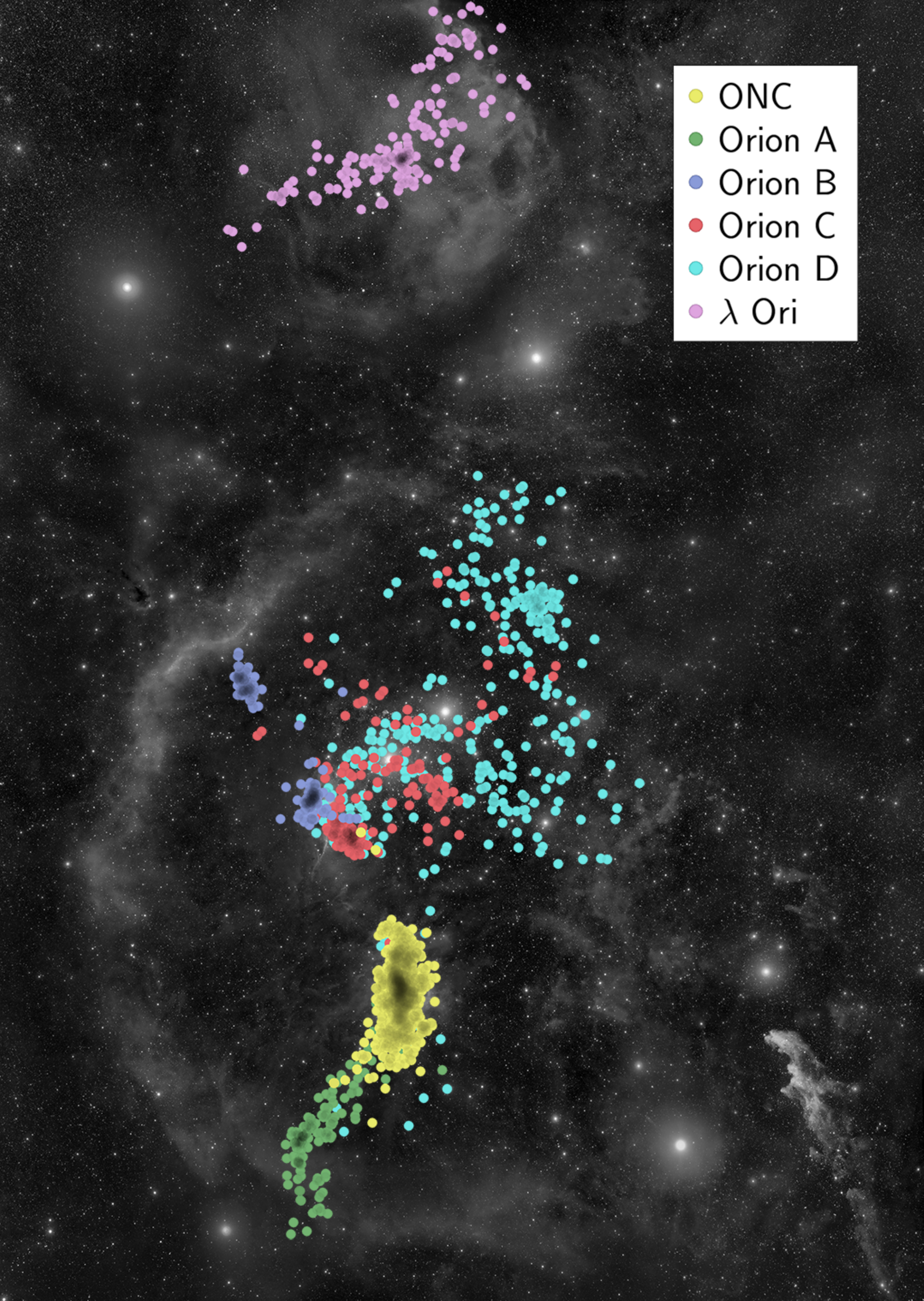}
    \caption{Foreground: each dot represents the projected position of a star in the Orion Complex that we include in our analysis. The star's color corresponds to its main group. Background: greyscale optical image of the Orion Molecular Cloud Complex, courtesy of Rogelio Bernal Andreo.}
    \label{fig:orion}
\end{figure}

\subsection{Data Acquisition}\label{sec:data1}
We obtain the locations, parallax distances, and proper motions of stars in the Orion Molecular Cloud Complex from Gaia's second data release \citep{Gaia18}. The Gaia catalog includes five-parameter astrometric data from over 1.3 billion stars. 
The line-of-sight velocities of the stars were observed in the near infrared during the Apache Point Observatory Galactic Evolution Experiment 2 (APOGEE-2). The APOGEE spectrograph is mounted on the 2.5 m Sloan Foundation Telescope of the Sloan Digital Sky Survey (SDSS), and is used to collect high resolution spectral images of over 400,000 stars in the Milky Way \citep{Gunn06,Blanton17}. 
In total, 8891 stars from APOGEE-2, and 16754 stars from Gaia DR2 were observed towards the Orion Complex, as specified by \citet{Kounkel18}. Combining these two surveys provides six-dimensional information of a subset of stars.

\citet{Kounkel18} developed a hierarchical clustering algorithm to assess membership of the Orion Complex and its structure based on the position of stars in the phase space (position+velocity). They identify five main groups inside the Complex, namely Orion A, Orion B, Orion C, Orion D, and $\lambda$ Ori. We require all stars in our sample to be a part of one of these groups, excluding all of the stars that could not be clustered together (i.e., field stars, or other members of Orion that may have peculiar velocity - such as spectroscopic binaries or runaways). Furthermore, we require a complete information on six-dimensional position and velocity for each star, i.e, a star needs to have both astrometric solutions from Gaia DR2 as well as radial velocities from APOGEE. Through this, we obtain a sample of 1439 stars.

We note that the Orion Nebula Cluster (ONC) is a part of the Orion A molecular cloud, however, it is very unique, being a singularly most massive young cluster in the solar neighborhood. The environment that is found in the ONC is different from the more diffuse environment in the southern part of Orion A. Thus, for the purpose of our analysis, we separate ONC and the rest of the Orion A. Throughout the paper, Orion A refers only to the ``tail'' of the cloud (L1641 and L1647), excluding stars in ONC. The reasoning behind this split is discussed in Section~\ref{sec:results}. The term ``group'' in this paper refers to all ``stellar systems'' that are spatially and kinematically associated, while ``cluster'' only refers to open clusters such as the ONC. 

Table~\ref{table:data} lists the number of stars and the mean age of each main group and the ONC. We plot the projected positions of the stars in our analysis over a greyscale image of the Orion Complex in Figure~\ref{fig:orion}.

\subsection{Method of analysis}\label{sec:method}
\begin{figure*}
    \centering
    \includegraphics[width=0.98\linewidth]{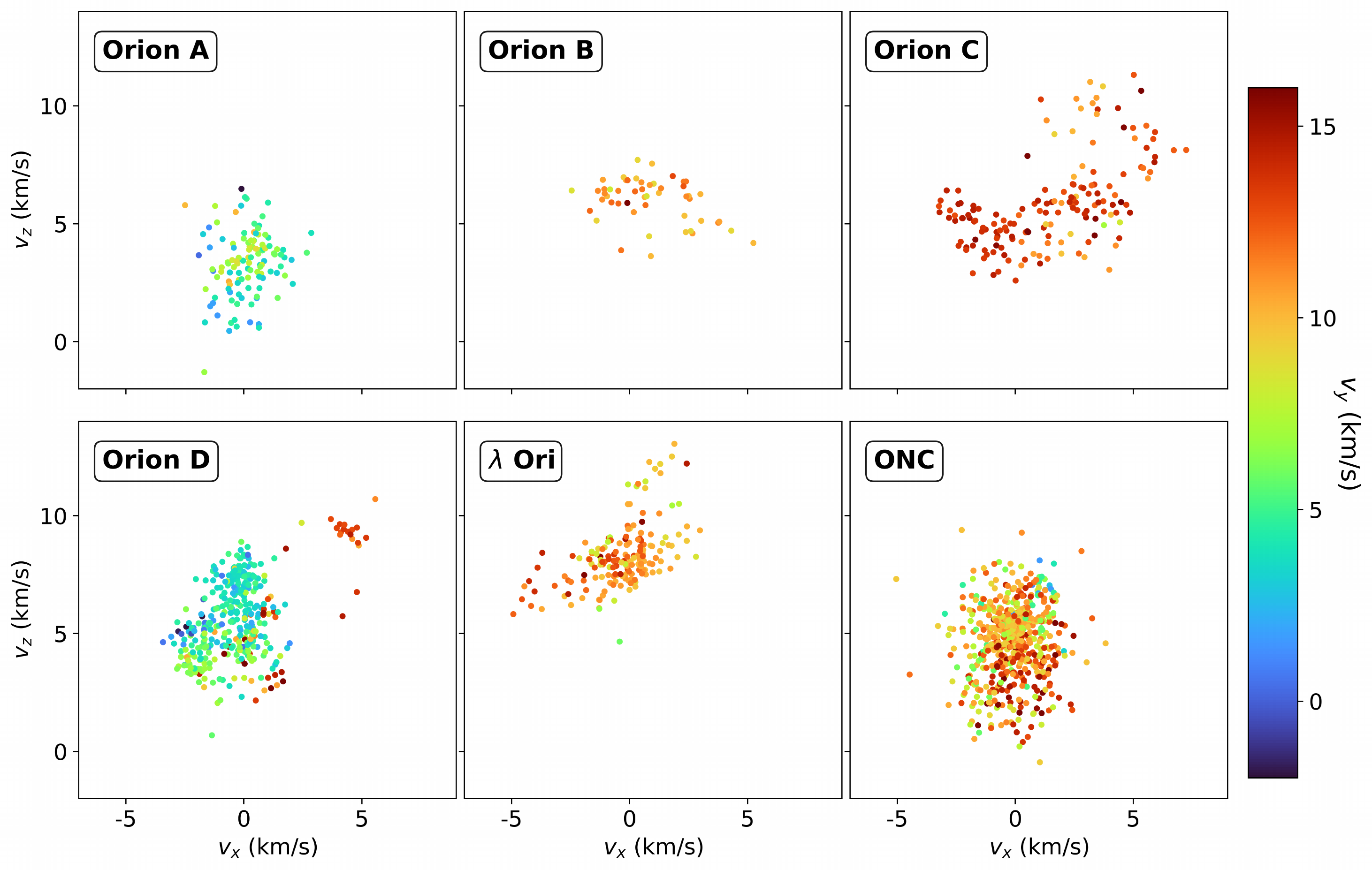}
    \caption{Velocities of stars in the six clouds in our analysis in Cartesian coordinate. The color map denotes stars' velocity in the y direction, which closely resembles their radial velocity. All clouds show some level of randomness in the velocities of the stars.}
    \label{fig:velodist}
\end{figure*}
To test our hypothesis that young star clusters ``remember" their local gas kinematics at the time of birth, we compute the VSFs for all identified groups, as well as the VSF of the combined Orion Complex catalog, both with and without the ONC. We first convert the radial velocity and proper motion of stars into velocity in Cartesian coordinate, in the Local Standard of Rest (LSR). Figure~\ref{fig:velodist} shows the 3D velocity information of stars used in this study in each cloud. 

For each cloud, we compute the first-order VSF and its uncertainties. The VSF for a sample of stars can be calculated in the following way: for each pair of stars, we first record the physical separation $\ell$ of the pair and compute the magnitude of the vector velocity difference $\delta \Vec{v}$ as the velocity difference $|\delta v|$. We then compute the average of the velocity differences $\langle |\delta v| \rangle$ within logarithmic bins of $\ell$. 

The VSF is related to the kinetic energy power spectrum of turbulence. It describes how velocity difference ($|\delta v_{ij}|$) relates to the physical separation ($\ell_{ij}$) for given pairs of points in space ($i$, $j$). In a turbulent flow, kinetic energy cascades down from large scales to small scales, and the velocity differences become smaller toward smaller scales. If the turbulence is in-compressible (subsonic), we expect the first-order VSF to be $|\delta v_{ij}|\sim \ell_{ij}^{1/3}$ within the inertial range \citep{Kolmogorov41}. For compressible (supersonic) turbulence, $|\delta v_{ij}|\sim \ell_{ij}^{1/2}$ \citep[e.g.,][]{Federrath13}.

To estimate uncertainties of the VSF, we first perform random sampling of the measurements of stars' properties based on the observed uncertainties, and obtain 1000 realizations for each star following a Gaussian distribution. We only consider measurement uncertainties of the parallax distances, proper motions, and radial velocities. We ignore the uncertainties in RA and Dec because they are typically much smaller than the uncertainties of other quantities. For each group, we then obtain 1000 VSFs, with each iteration excluding a random star within the group. This allows us to account for both measurement errors and uncertainties due to small sampling sizes in some bins of $\ell$. We take the mean and standard deviation of the computed VSFs to be the group's VSF and its corresponding error. We choose 1000 iterations to ensure accuracy even though the results are already converged at 100 iterations.

We also compute the second and third-order VSFs of the groups, and discuss them in the Appendix.

\begin{table*}[ht]
\begin{center}
\caption{Properties of molecular clouds in the Orion Complex}\label{table:data}
\hspace*{-3cm}
\begin{tabular}{|c|c|c|c|c|c|c|}
\cline{1-7}

   & Orion A & Orion B & Orion C & Orion D & Lambda Ori & ONC \\
\cline{1-7}
Number of Stars$^a$ & 117 & 51 & 162 & 354 & 170 & 585  \\ 
\cline{1-7}
Median Age (Myr) & 2.2 & 1.3 & 4.1 & 4.9 & 3.7 & 2.7  \\ 
\cline{1-7}
Crossing time (Myr) & 16.1 & 13.6 & 8.0 & 12.9 & 12.8 & 6.9$^b$  \\
\cline{1-7}
Fitting Range$^c$ (pc) & 10 - 90 & 8 - 60 & 9 - 50 & 15 - 58 & 10 - 32 & 5 - 40  \\
\cline{1-7}
Slope of VSF & $0.18$  & $0.14$ & $0.19$ & $0.53$ & $0.36$ & $0.05$  \\
\cline{1-7}
\end{tabular}
\\
\vskip 0.1in 
\raggedright{Note. a. Number of stars obtained from the combined Gaia DR2 and APOGEE surveys with full 6D information (Section~\ref{sec:data1}).
b. This is the crossing time for the entire ONC cluster. The inner core has a much shorter crossing time and relaxation time \citep{Hillenbrand98}. c. The slope of the VSFs depends on the exact range of separation taken into account. We choose the fitting range such that the uncertainties are low and there is a consistent trend in the slope of the VSF.} 
\end{center}
\end{table*}

\begin{figure*}
\centering
\includegraphics[width=0.32\linewidth]{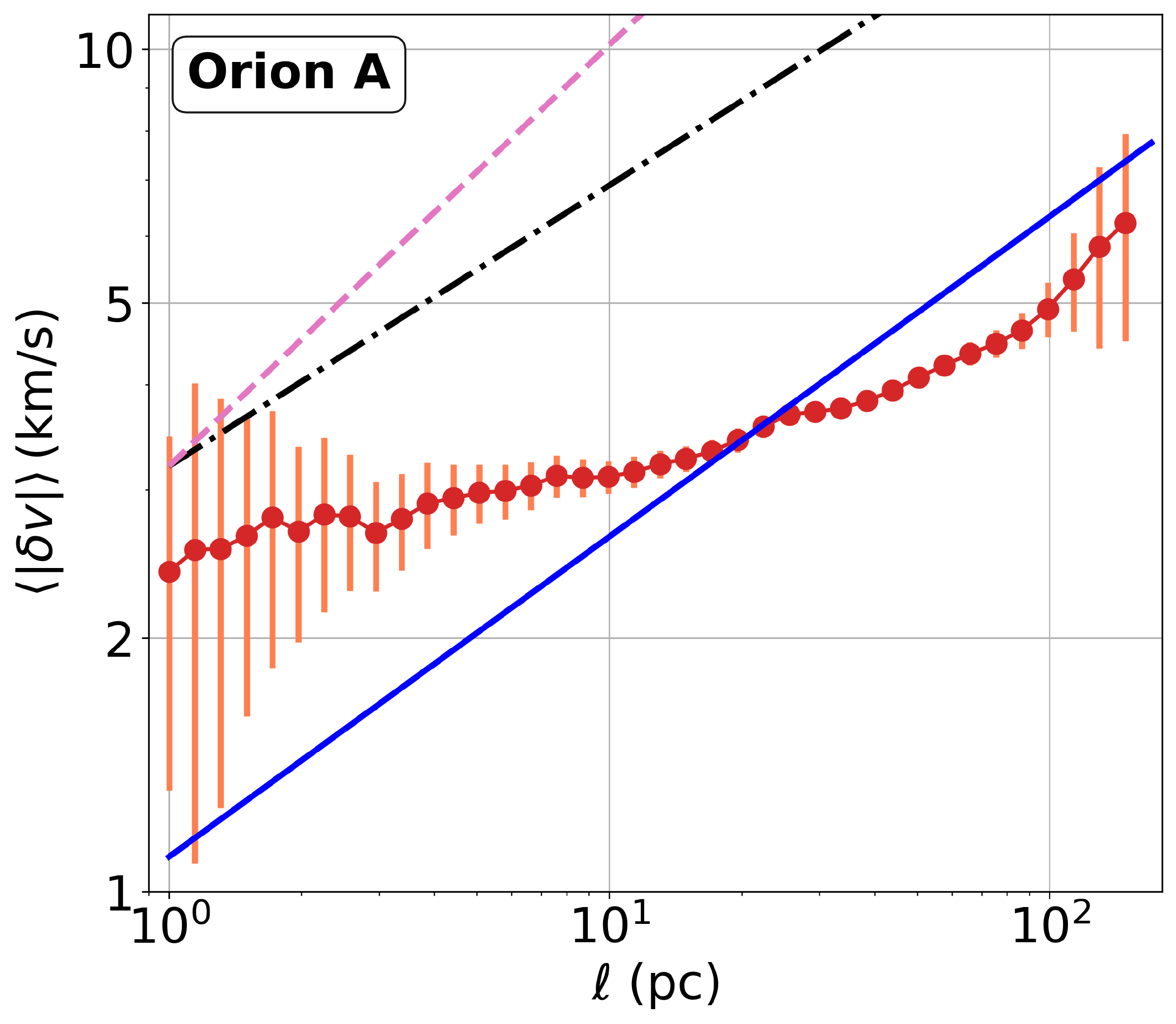}
\includegraphics[width=0.32\linewidth]{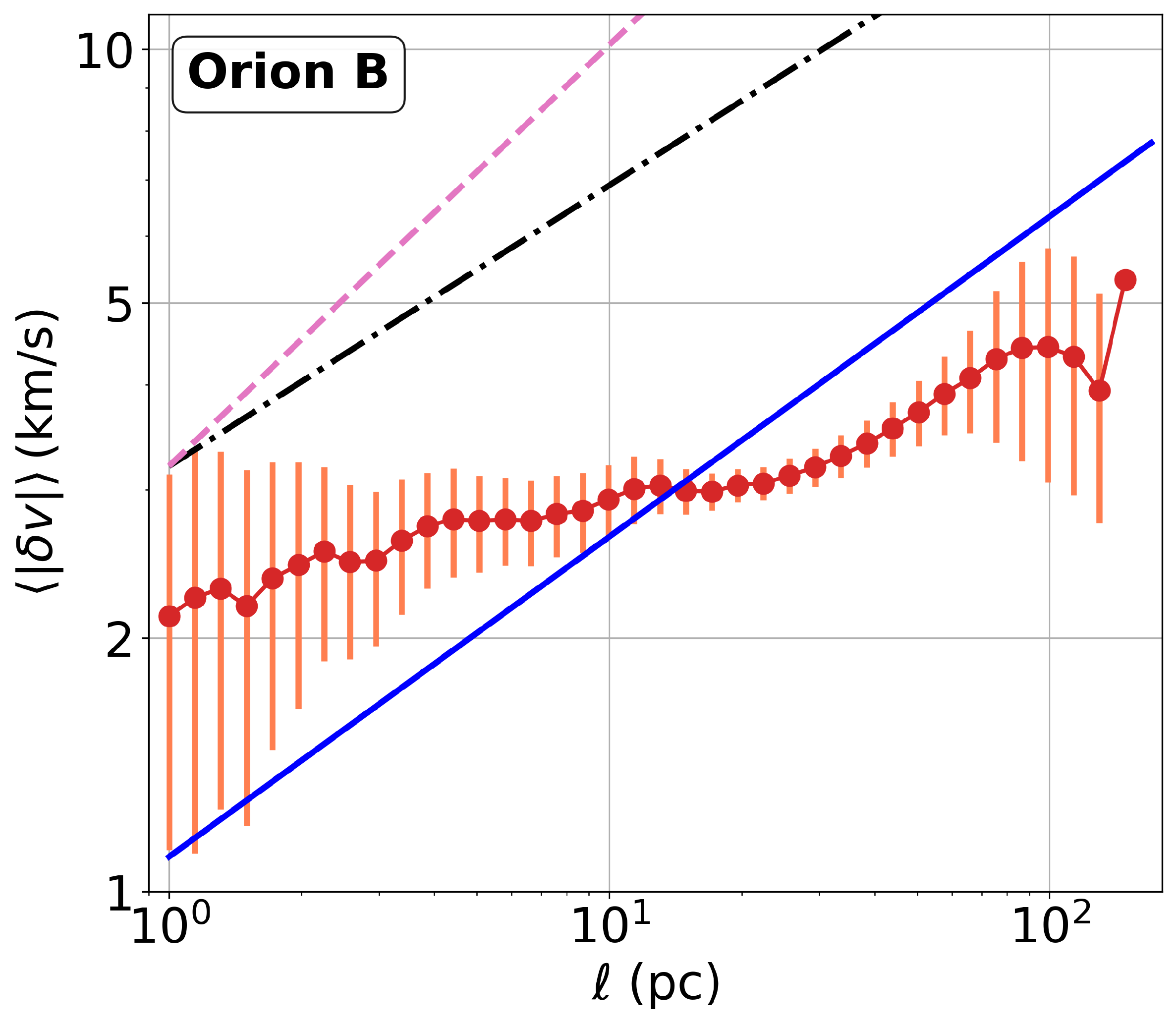}
\includegraphics[width=0.32\linewidth]{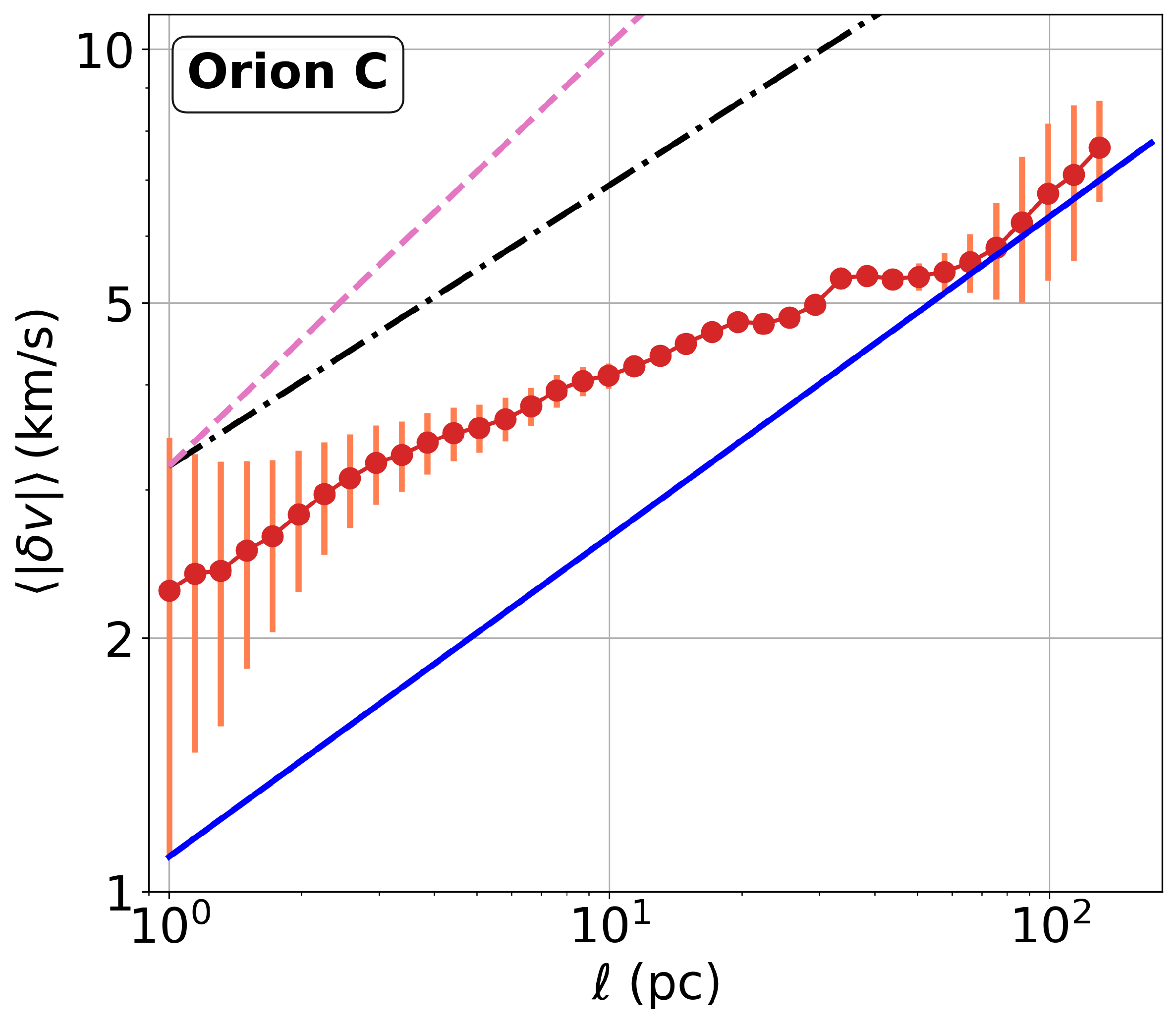}\\
\vskip 0.1in
\includegraphics[width=0.32\linewidth]{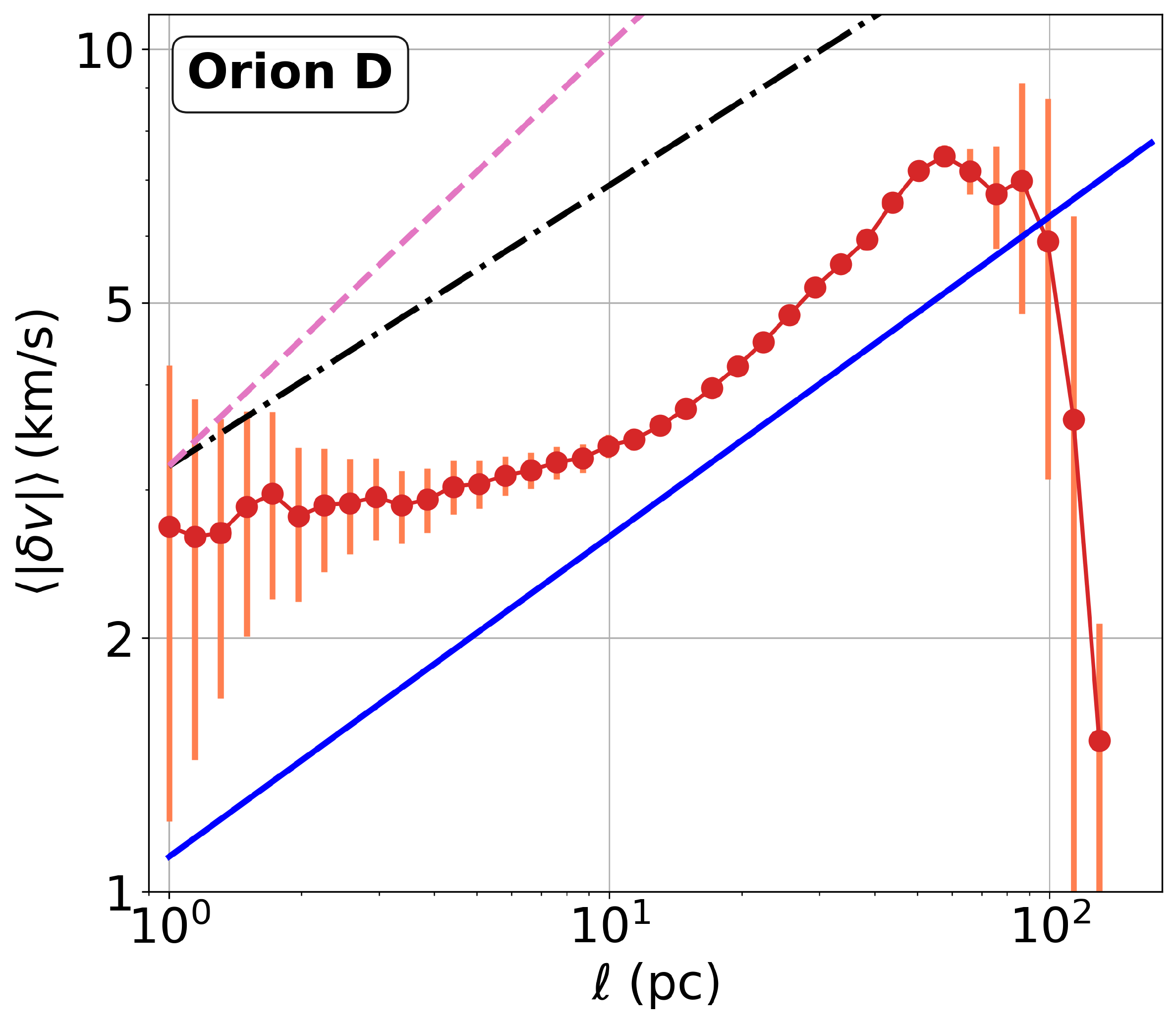}
\includegraphics[width=0.32\linewidth]{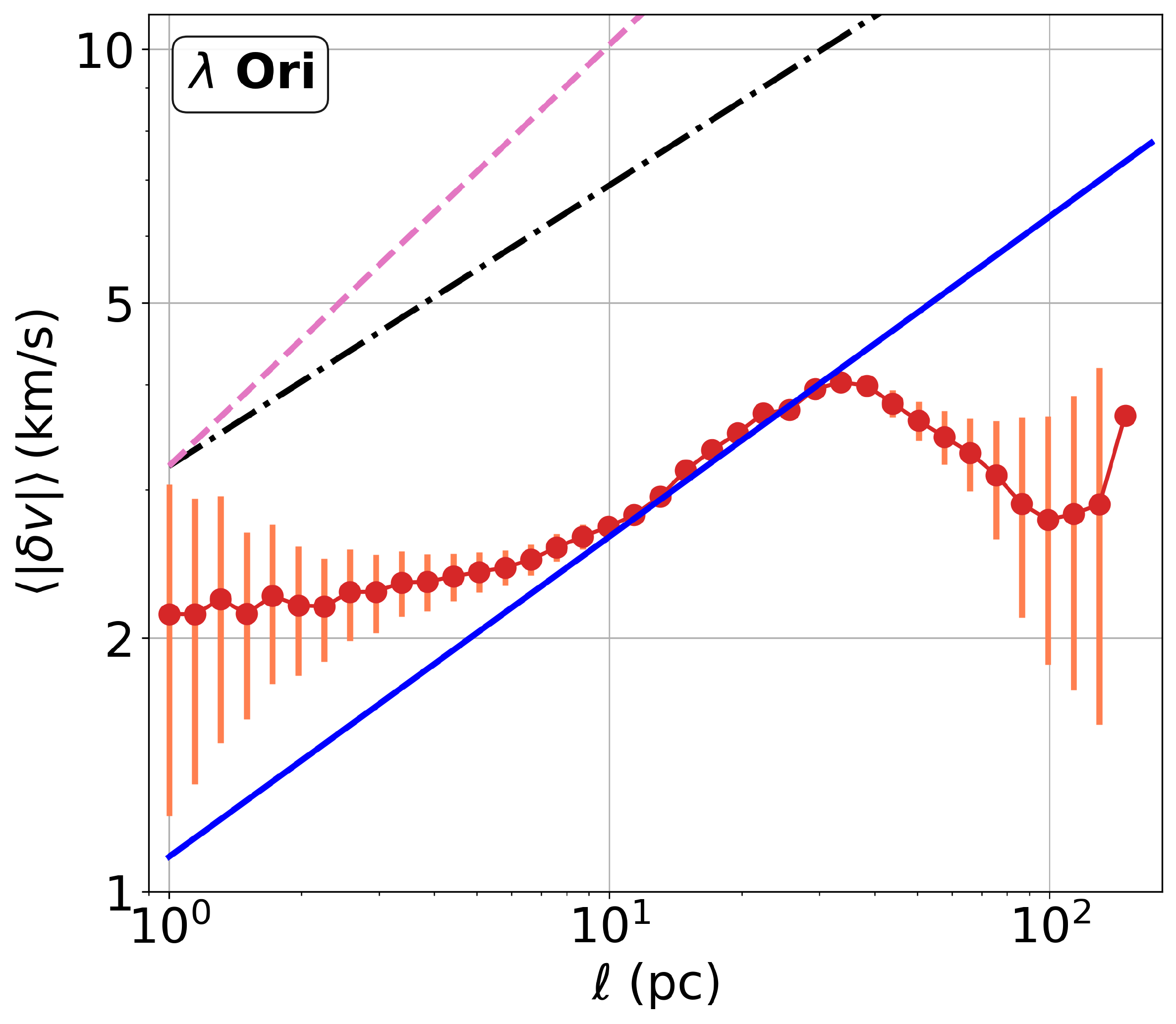}
\includegraphics[width=0.32\linewidth]{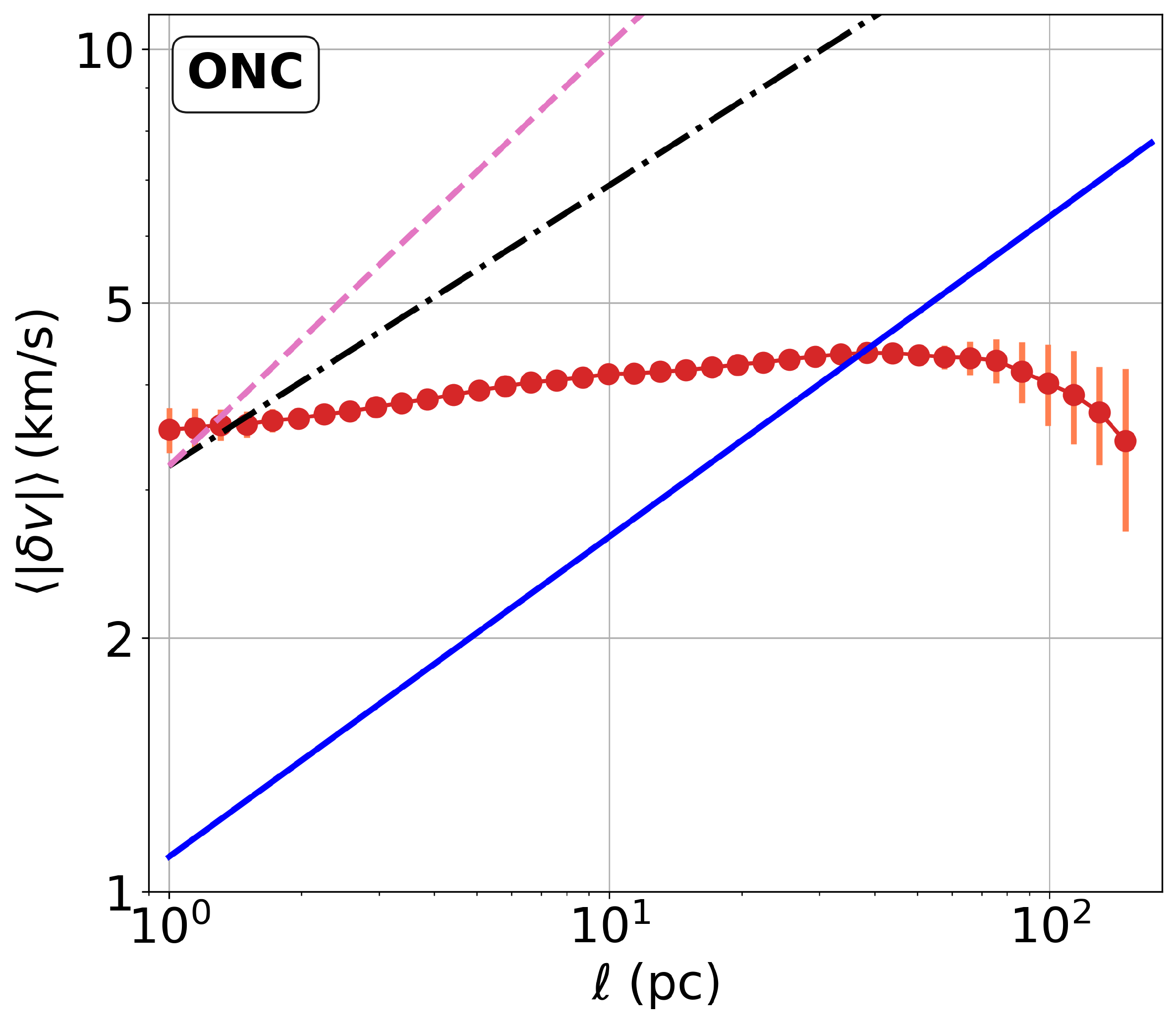}\\
\vskip 0.1in
\includegraphics[width=0.32\linewidth]{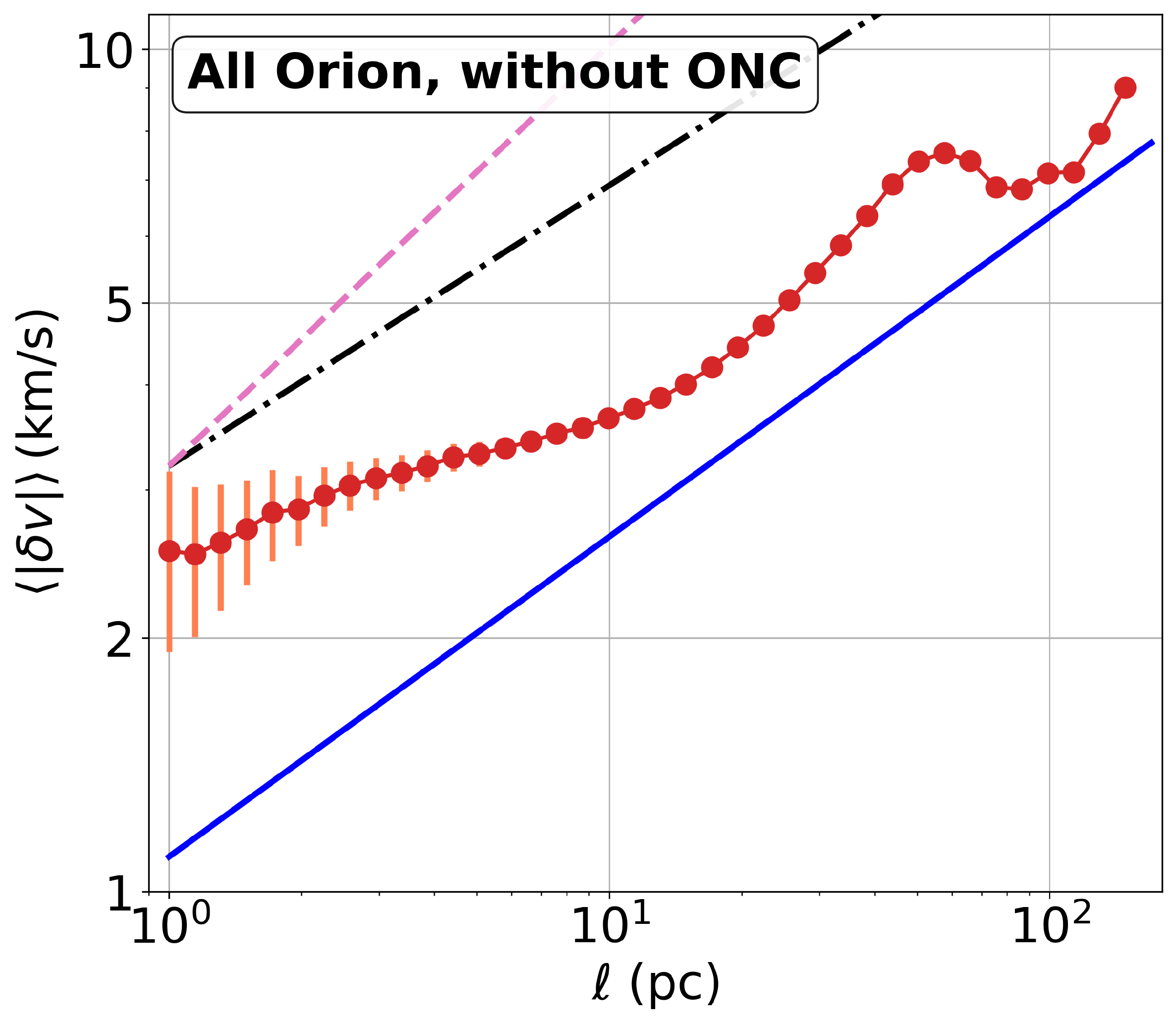}
\includegraphics[width=0.32\linewidth]{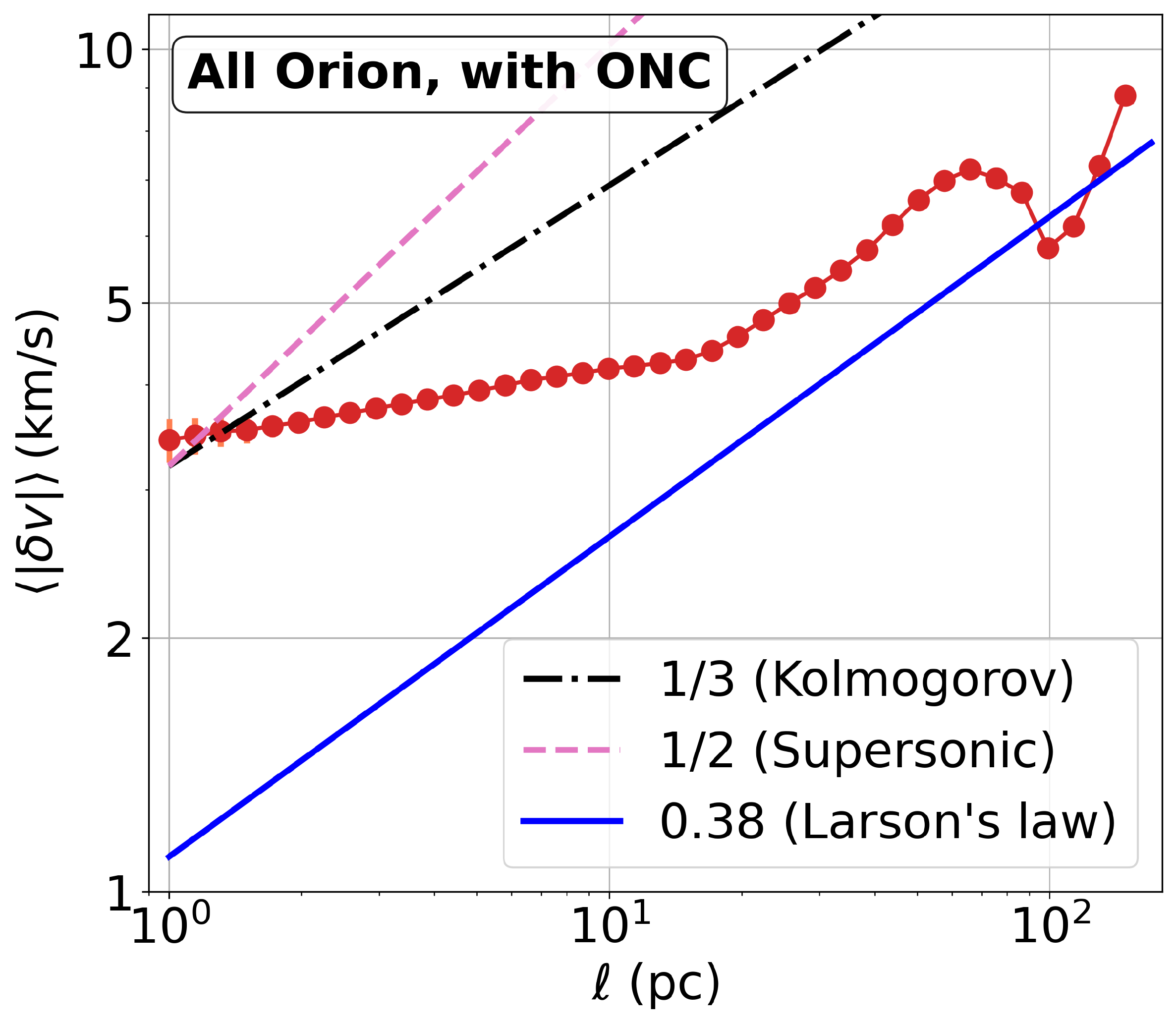}
\caption{Left to right, top to bottom: first-order VSF of stars in the Orion A, Orion B, Orion C, Orion D, Lambda Ori, ONC, all Orion without ONC, and all Orion with ONC. The error bars are obtained through a random sampling analysis, and represent the uncertainties from both observational errors and uncertainties due to small sample size. As reference, we also plot a black dot-dashed line of slope 1/3 for the Kolmogorov turbulence, a pink dashed line of slope 1/2 for supersonic turbulence, and a solid blue line for Larson's velocity dispersion correlation for molecular clouds: $\sigma (km \cdot s^{-1}) = 1.1 \cdot L (pc)^{0.38} $ \citep{Larson81}. All panels use 40 logarithmic bins of $\ell$ from 1 pc to 170 pc. }\label{fig:vsf}
\end{figure*}

\section{Results}\label{sec:results}

Figure~\ref{fig:vsf} shows the first-order VSF of stars in all six individual groups in the Orion Complex. The last two panels show the combined VSF of all groups without and with the ONC, respectively. 
To guide the eye, we also plot in each panel the expected scaling for supersonic and subsonic turbulence, along with Larson's law measured for turbulence in molecular clouds in the Milky Way \citep{Larson81}.

The VSFs of all six individual groups except ONC exhibit a power-law scaling, characteristic of a turbulent flow. The amplitudes and slopes vary from group to group. The best-fit VSF slopes are listed in Table~\ref{table:data}.

The Orion A group (top left panel of Figure~\ref{fig:vsf}) is located just outside of Barnard's Loop (the shell-like feature to the left of Orion A-D in Figure~\ref{fig:orion}, see also \citet{2015ApJ...808..111O}), which is proposed to have originated from a supernova that erupted $\sim$6 Myr ago \citep{Kounkel20}. Due to its location, Orion A is largely unaffected by the supernova's shock wave. Indeed we observe that its VSF slope generally follows that of subsonic turbulence. The overall amplitude is slightly below Larson's relation. 

The Orion B cloud is connected to Orion A but closer to the center of the supernova explosion. Our analysis of the Orion B group suffers from high uncertainties due to low number of stars (51) available in the catalog. The slope of the VSF of Orion B (top middle panel) is not well constrained, but the VSF generally follows the turbulence-like kinematics present in other clouds. The amplitude is roughly consistent with Larson's relation.

Orion C lies within the supernova bubble, and as \citet{Kounkel20} suggested, was driven apart from Orion D at the opposite front of the supernova shock wave. Most of the stars in Orion C and D formed after the supernova explosion that created the Barnard's Loop. We observe this local energy injection effect on the VSF of Orion C (top right panel) with a higher-than-Larson amplitude. 
The slope of the VSF is flatter than the Kolmogorov turbulence due to the presence of $\sigma$ Ori, a dense cluster of age $\sim$2 Myr. We discuss the effects of dense clusters on the VSF in more detail later in this section. 

Stars in the Orion D group exhibit a curious VSF (middle left panel), with a slope more in line with supersonic turbulence (slope $1/2$) than Kolmogorov at distance scale $10<\ell<60$ pc and absolute velocity difference higher than that of Larson's law. We attribute these properties of Orion D's VSF to its position inside the supernova bubble. Located close by and at the front of the bubble (in our line-of-sight) as it expands, motion of stars in the Orion D cloud is additionally driven by the shock wave of the supernova, hence its turbulent kinetic energy is amplified.

The last independent group we look at is $\lambda$ Ori (center panel of Figure~\ref{fig:vsf}). In the $10<\ell<32$ pc range, the VSF of stars in $\lambda$ Ori closely exhibits the 0.38 slope of Larson's law. The amplitude also exactly follows Larson's relation \citep{Larson81}, and is higher than Orion A and B, but lower than C and D. This result can be understood given $\lambda$ Ori's history and isolated position in the Orion Complex. The $\lambda$ Orion bubble (the bubble-like structure surrounding $\lambda$ Ori in Figure~\ref{fig:orion}) is proposed to have been created by a supernova explosion $\sim1$ Myr ago \citep{Dolan2002}. Although \citet{Kounkel20} estimated the age of the remnant to be $\sim 4$ Myr. Regardless, the explosion happened after many of the stars have already formed in $\lambda$ Ori where the stellar age has a wide spread over $\sim 6$ Myr \citep{Kounkel18}. Thus the effect of local energy injection is not as prominent as in Orion C and D.

The ONC's VSF (middle right panel) has a distinctively flatter slope compared to other groups in the Orion Complex. The ONC is a dense star cluster located at one end of the Orion A molecular cloud (Figure~\ref{fig:orion}). Unlike the loose star groups discussed previously, the ONC has a dense inner core ($\sim 2$ pc) with an estimated relaxation time of $\tau_r \sim$ 0.6 Myr \citep{Hillenbrand98}, shorter than the cluster's mean age of $\sim$ 2.6 Myr. Therefore, a significant fraction of the stars in the ONC have dynamically relaxed, and have hence ``forgotten" the turbulence kinematics of their natal clouds. The residual non-zero slope likely originates from the stars in the outer parts of the cluster that have not relaxed yet. It is worth noting that if we simply look at the velocity distributions of stars in different groups (Figure~\ref{fig:velodist}), ONC does not appear drastically different from the rest of the groups apart from its compactness in the velocity space. Yet our analysis reveals its distinct kinematics due to relaxation.

In the last two panels of Figure~\ref{fig:vsf}, we compute the VSF for all the stars within the Orion Complex without and with the ONC (bottom left and bottom right panels, respectively). Without the ONC, the VSF of the Orion Complex has similar characteristics to that of Orion D, with a steeper-than-Kolmogorov slope and an amplitude higher than Larson's law in the $10<\ell<60$ pc scale range. This is because D has the highest number of stars amongst the 5 groups (Table ~\ref{table:data}). 

When the ONC is included, on small scales ($2<\ell<20$ pc), the VSF flattens because it is heavily influenced by the cluster due to its dominant number of stars in the catalog (585 stars vs. a total of 1439 stars). In fact, this is why we separate the ONC from Orion A in our analysis. If we include ONC in Orion A, the resulting VSF is significantly flatter. All the other star groups also contain star clusters. For example, Orion C has $\sigma$ Ori. We do not separate these clusters because they are much smaller than the ONC, and removing them degrades the statistics of our analysis. We note, however, that they may have contributed to the flatter VSF of Orion C and the flattening of the VSF in the other groups on small scales. We discuss other uncertainties in more detail in Section~\ref{sec:discussions}. 

The combined VSF shows a bump at $\sim 60$ pc, similar to Orion D, suggesting energy injection at that scale. The mean age of the stars in this study is $\sim 3.7$ Myr. Given the current size of the Barnard's loop ($\sim 80$ pc) and its age ($\sim 6$ Myr), the energy injection scale is roughly consistent with the size of the supernova bubble when the stars first formed. This again shows the effects of local energy injection of supernova explosions on turbulence.

\section{Discussions}\label{sec:discussions}

\subsection{Biases and Uncertainties}

Gas in molecular clouds such as the Orion Complex is turbulent by nature. The theoretical foundation of our analysis is that newborn stars retain this turbulence information. However, after the stars are born, they decouple from the hydro-dynamical forces of the gas, and no longer follow the turbulent flow. In other words, stars are only perfect tracers of turbulence at birth. Since star formation happens over a period of time, all of the star groups contain stars that are over a few Myr old. We have identified two mechanisms through which stars can lose their memory of the ISM turbulence, causing biases in our analysis.

The first mechanism is dynamical relaxation, which we discussed in the analysis of the ONC in Section~\ref{sec:results}. Due to open clusters' high density, close stellar encounters are common, which alter the stars' original orbits and lead to dynamical relaxation. The relaxation time is estimated as a function of the number of stars and the crossing time of the cluster:

\begin{equation}
t_{relax} = t_{cross} \cdot \frac{N}{6 \cdot \ln{N/2}} \,,   
\end{equation}
Where N is the total number of stars in the cluster and $t_{cross}$ is the crossing time, defined as the typical time for a star to travel a distance equal to the half-mass size of the cluster:

\begin{equation}
t_{cross} =  \frac{2 \cdot R_{hm}}{\sigma_v} \,.
\end{equation}
Here, $R_{hm}$ is the radius within which half of the cluster's stars reside, and $\sigma_v$ is the bulk-motion-removed characteristic velocity of the cluster's stars. It is computed as the standard deviation of the velocity distribution with respect to the median \(\sigma_v \equiv \sigma [|v_i - v_{median}|]\).

After the stars are relaxed, they no longer hold the turbulent kinematics of gas at the time of their formation. For loose groups of stars such as those we analyze, the crossing times are significantly longer than the group's current age (Table~\ref{table:data}). Furthermore, since each group possesses a number of stars in the order of N $\sim 10^3$ (eg. N = 2800 stars within the dense core of the ONC was used by \citet{Hillenbrand98}), the estimated relaxation time is an order of magnitude longer than the crossing time. Therefore, dynamical relaxation has most likely not affected the shape of the VSF for the five groups in the Orion Complex, except for the ONC. This is in agreement with \citet{DaRio2017} who find that young stars in Orion A are kinematically associated with the molecular gas but the ONC appears more dynamically evolved and virialized. \citet{Kounkel18} also found consistent kinematics between young stars and gas in the whole Orion Complex where the gas has not been dispersed. 

In addition to the ONC, there are other clusters within each group: $\sigma$ Ori in Orion C, 25 Ori in Orion D, $\lambda$ Orionis Cluster in $\lambda$ Ori, and two small clusters NGC 2024 and NGC 2068 in the Orion B group. These clusters have much smaller physical sizes than the ONC and contain only a small fraction of stars of the group. This property of the clusters can be visually confirmed from Figure~\ref{fig:orion}. Thus, their existence does not significantly affect our overall results. However, they may have contributed to the flattening of the VSFs on small scales that we observe in almost all individual groups.

Another potential source of bias in our analysis comes simply from the drifting of stars from their original positions. Assuming no strong stellar encounters, stars with higher velocity differences drift apart at a faster rate than stars with smaller velocity differences, causing the VSF to steepen with time. This steepening happens over roughly $t_{cross}$.

In our analysis, most groups are relatively young, with their ages much shorter than $t_{cross}$ (Table~\ref{table:data}), so drifting should not have had time to erase turbulence-driven velocity differences. Although drifting may have steepened the VSF of Orion D, which has the oldest median age and the steepest slope amongst all the groups analyzed here. We plan to further explore the effects of relaxation and drifting in our future works both with a larger observed sample and with numerical simulations \citep{Hui2019}.

There are also other sources of uncertainties in our analysis, such as binary systems. While the barycenter should follow the turbulent velocity of the natal cloud, binaries also have orbital motion contributing to the overall velocity measurement. While most systems with very large orbital speeds have been excluded from the analysis through the initial hierarchical clustering, systems with orbital speeds of only a few km s$^{-1}$ likely remain.

False positives in hierarchical clustering, such as contamination from the field stars can also affect our results. \citet{Kounkel18} estimated that false positive fraction for the clustering algorithm can be as high as $\sim$6\%. In addition, drifting can result in close pairs of uncorrelated stars with high relative velocity. Binaries, drifting, and mis-grouped stars, coupled with small sample sizes at small separations would explain the flattening of the VSFs at small $\ell$.

\subsection{Comparison with other Methods to Measure Turbulence}
Most of the previous studies on turbulence in astrophysical environments rely on the observations of gas. Although only velocity statistics can directly 
reflect the dynamics of turbulence and turbulent energy cascade, 
it is the density statistics that is widely used for measuring turbulence as it is much more easily accessible to observations. For example, the delta-variance method has been used to analyze the structure of observed molecular cloud images in the ISM \citep{Stutzki1998}. The same method and a modification of it \citep{Mexican2012} have also been used to infer turbulence in the hot intra-cluster medium based on X-ray images \citep[e.g.,][]{Zhuravleva2014}.

Velocity centroids can recover the turbulent velocity spectra in subsonic turbulence 
\citep{Esq05}. 
The velocity channel analysis (VCA)
using channel maps and 
velocity coordinate spectrum (VCS)
using the fluctuations 
measured along the velocity axis of the Position–Position Velocity (PPV) cubes
can also be used to extract turbulent velocity spectra in supersonic turbulence
(see \cite{Laz09} for a review). 
In particular, the VCS technique only requires measurements along a few directions to obtain a reliable velocity spectrum
\citep{CheL09}. Spectroscopic data has also been used in the Principal Component Analysis to extract the size-line width relation and recover the structure function within molecular clouds \citep[e.g.,][]{Heyer04,Roman-Duval11}.

Recently, statistical measurements of the velocities of gas phase point sources have been developed for, e.g., dense molecular cores embedded in the background turbulent flow \citep{Qi12} and filaments in the centers of galaxy clusters \citep{Li20}. These analyses use the 1D line-of-sight velocities and 2D positions projected on the plane of the sky. The bias due to the projection effect depends on the thickness of the structure \citep{Qi15}, which is usually unknown.

Our analysis in spirit is similar to that of gas phase point sources, but instead of using projected information, we take advantage of the full 6D (3D velocity + 3D position) information of stars. Thus our results do not suffer from the poorly constrained projection uncertainties. Observing 6D information of the gas, especially the motions in the plane of the sky, is prohibitively challenging \citep[although see][]{2008AJ....136.1566O}. Thus our method has a unique advantage, and can potentially be used to help better understand the biases in the analyses based on gas.

\subsection{Comparison with Theories and Other Observations}

There has been a lot of previous work on turbulence in Orion based on the observations of emission lines of the ionized gas \citep{1988ApJS...67...93C, 1992ApJ...387..229O, McLeod2016, 2016MNRAS.463.2864A}. Most of them are focused on small scales ($<1$ pc) and therefore can not be directly compared with our results. But they do generally find VSFs consistent with turbulence, in agreement with our results. 

On larger scales (a few to $\sim 100$ pc), we have compared our results with Larson's Law in Section~\ref{sec:results}. Individual groups may show slightly higher or lower amplitude than the mean Larson's relation, with slightly steeper or shallower slopes, but overall, our results are in good agreement with Larson's Law for molecular clouds in the Milky Way. Notably, \citet{Larson81} also finds a slightly lower-than-average amplitude for Orion B, consistent with our findings here even though we are using different tracers of turbulence. 

Our results show that regions heavily influenced by supernova explosions can exhibit a higher level of turbulence, and even a steeper VSF. This is consistent with numerical simulations showing that supernovae can drive turbulence in molecular clouds \citep{Pad16}. 
Additionally, the steepened slope of Orion D is consistent with \citet{Federrath13} finding that supernova explosions excite more compressible than solenoidal velocity modes in the gas, which tends to steepen the VSFs, and thus exhibit the 1/2 power law of supersonic turbulence. Observations have also shown the local effects of supernovae on turbulence in other systems (e.g., in the Large Magellanic Clouds \citep{2019ApJ...887..111S}). 

\section{Concluding Remarks and Future Work}
We have introduced a new method to study turbulence in the ISM using motions of 
young stars. Our results demonstrate that stars in young star groups retain the memory of turbulence of their natal molecular clouds. Our analysis uses the full 6D information of stars to trace turbulence, and does not suffer from projection uncertainties. Our results can be used to provide independent constraints on the turbulent properties of molecular clouds in addition to analysis of gas kinematics, and also shed light on studies of the dynamical evolution of star clusters in the Galaxy \citep{Kam19}. 

We plan to apply our analysis to a much larger sample of star groups in the future. Meanwhile, we will perform and analyze tailored numerical simulations of star formation in molecular clouds to better understand biases and uncertainties in our analysis, as well as constraining subgrid models for star formation \citep{Hui2019}.

\section*{Acknowledgments}
We would like to thank Anna McLeod and Yuan-Sen Ting for helpful discussions. This work was partly performed at the Aspen Center for Physics, which is supported by National Science Foundation grant PHY-1607611. HL is supported by NASA through the NASA Hubble Fellowship grant HST-HF2-51438.001-A awarded by the Space Telescope Science Institute, which is operated by the Association of Universities for Research in Astronomy, Incorporated, under NASA contract NAS5-26555.
S.X. acknowledges the support provided by NASA through the NASA Hubble Fellowship grant 
$\#$ HST-HF2-51473.001-A awarded by the Space Telescope Science Institute, which is operated by the Association of Universities for Research in Astronomy, Incorporated, under NASA
contract NAS5-26555.

\begin{figure}
    \centering
    \includegraphics[width=0.7\linewidth]{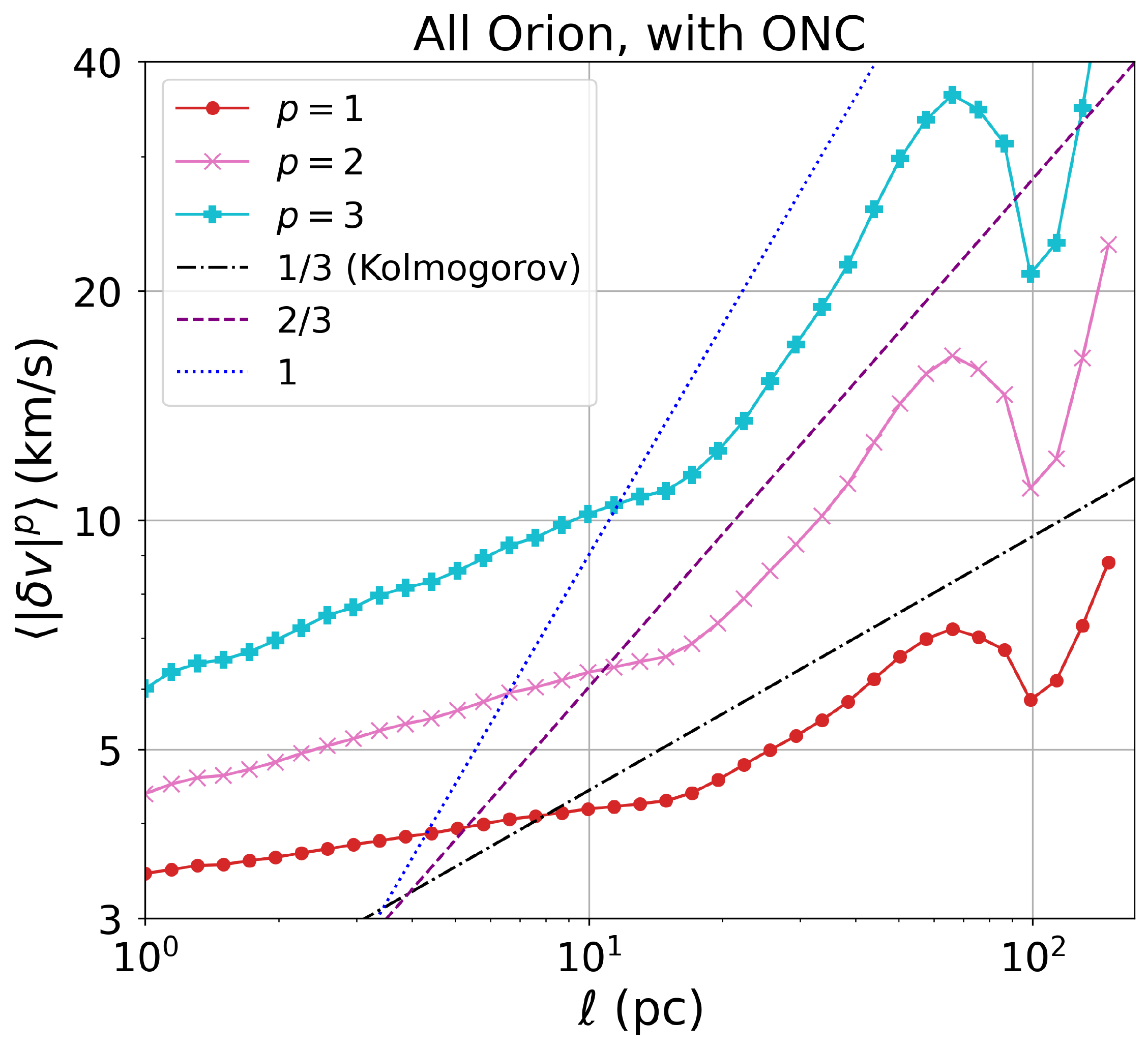}
    \caption{First ($p=1$, red), second ($p=2$, pink), and third ($p=3$, blue) order VSFs of the stars in the entire Orion Complex. The second and third order VSFs have been shifted lower for easy comparison.}
    \label{fig:3rdorder}
\end{figure}

\appendix 
We present the second and third order VSFs in Figure~\ref{fig:3rdorder}. 
High order VSFs are usually used to evaluate the intermittency 
\citep{She94,Bold02}. 
Different behaviors of the high-order VSFs measured with different gas tracers, e.g., $H_2$, $CO$, were found in numerical simulations of molecular clouds,
which are related to their different volume filling factors and spatial distribution 
\citep{Bert15}. 
The second and third order VSFs that we present in Figure~\ref{fig:3rdorder}
are generally consistent with the Kolmogorov scaling. This finding agrees with the 
measurement using the velocities of dense cores in Taurus cloud
\citep{Qi18}. 
It suggests that both the dense cores and the stars used here sample the turbulence in the relatively diffuse region with a large volume filling factor, and thus the Kolmogorov scaling is expected \citep{Xu20}.

\label{sec:2ndvsf}


\bibliographystyle{aasjournal}

\end{document}